# Adaptive Optimization of Autonomous Vehicle Computational Resources for Performance and Energy Improvement

Saurabh Jambotkar, Longxiang Guo and Yunyi Jia

*Abstract*— Autonomous vehicles usually consume a large amount of computational power for their operations, especially for the tasks of sensing and perception with artificial intelligence algorithms. Such a computation may not only cost a significant amount of energy but also cause performance issues when the onboard computational resources are limited. To address this issue, this paper proposes an adaptive optimization method to online allocate the onboard computational resources of an autonomous vehicle amongst multiple vehicular subsystems depending on the contexts of the situations that the vehicle is facing. Different autonomous driving scenarios were designed to validate the proposed approach and the results showed that it could help improve the overall performance and energy consumption of autonomous vehicles compared to existing computational arrangement.

## I. Introduction

A modern autonomous vehicle requires a large amount of computational power to run its systems. Subsystems such as chassis control, perception, motion planning, etc. require computational resources. Such resources can be bandwidth, processor cores, memory allocation, etc. By conventional methods, the allocation of these resources are fixed when required. Systems such as the visual perception module that requires a large amount of computational power are given preference by allocating a larger portion of available resources and the rest of the systems having lower requirements are given resources accordingly. However, such resource requirement changes over time. Limiting the resource allocated to a system considering only its overall usage rather than its instantaneous requirement puts restrictions on the maximum performance of the system.

Many existing research efforts have been focusing on the adaptive allocation of computational resources, most of which prioritize cloud computing or internet of things (IOT) related applications. A two-level resource management frameworks is used by Kephart in [1] by evaluating system utility function that consists of system inputs and performance measures. The system inputs are considered to be power consumption and performance of the CPU measured by the response time of the system. Considering the nature of the cloud computation framework, a priority-based assignment of resources seems intuitive. Such a study is given by [2] to maximize profit earned by a cloud server. An approach with a mathematical model of the system is explored in [3], [4]. The use of discrete nonlinear and linear models for calculating system response and job queuing time based on the frequency of the server is shown in these studies. By utilizing optimal control methods, the resource allocator optimizes the performance of a local system that may be further used by the co-operative resource allocation process over a wide network [5]–[8]. Zhan [9] demonstrated an approach by applying deep reinforcement learning for the dynamic resource allocation of federated learning to achieve better convergence speed and energy efficiency. Liang [10] used deep reinforcement learning to optimize the resource allocation in vehicular networks. Huang [11] and Sun [12] applied deep reinforcement learning to the autonomous resource slicing for virtualized vehicular networks.

While those existing cloud computing/IOT related methods can be applied to inter-vehicle allocation of computational resources, not many options can be found for in-vehicle resource allocation. [13] proposed resources allocation method for vehicular attentive vision system that emphasized on regions of interest of a visual system and the resource allocation for analyzing the visual information based on the criticality of objects, time to collision, and their severity classification. [14] employed a genetic algorithm-based resource allocation approach to lower the energy cost of vehicular applications without violating hard real-time constraints. [15] and [16] introduced a centralized architecture for in-vehicle computing and developed optimal resource scheduling method for the architecture.

Although the problem of managing resources for an autonomous vehicle seems similar to these studies have tackled, the challenges offered by it is still unsolved. First, system performance indices for an automotive are not easy to observe. For estimating the accuracy of a perception system, prior data is needed and models must be developed that take resources available to the system into consideration. To implement such a concept for onboard resource management, a large amount of training and testing is needed to obtain positive results. A priority-based approach is simple and quick but does not cover multi-objective goals such as safety, accuracy, and might not give the required robustness. The energy consideration for an automotive is more of a concern of total energy consumption rather than power demand. Unlike a cloud server, the computational load of an automotive may go down significantly depending on the scenario vehicle is facing. The resources used can be reduced

---

Saurabh Jambotkar, Longxiang Guo and Yunyi Jia are with the Department of Automotive Engineering, Clemson University, Greenville, SC 29607 USA. (e-mail: sjambot@g.clemson.edu; longxig@clemson.edu; yunyij@clemson.edu).

and lower resource consumption can result in lower energy consumption without loss of functional metrics [17].

Therefore, in this paper, an adaptive optimization approach is proposed to allocate onboard computational resources and an autonomous vehicle to different subsystems as the situation changes. For doing so, three performance metrics of the system are decided that are time, precision, and energy consumption. The subsystems are prioritized based on contextual information. Computational resources are allocated by optimizing system performance functions. Additionally, a safety metric is proposed and is evaluated in operational situations. The system performance is simulated in different scenarios and the results are compared with conventional resource distribution methods. The contributions of this paper can be summarized as follows: 1) create mathematical modeling of computational factors in autonomous vehicle; 2) propose adaptive computational resource optimization with context-driven subsystem prioritization; and 3) evaluate performance and design metrics for different autonomous driving scenarios.

## II. Adaptive Computation Optimization for Autonomous vehicles

### A. Problem Statement

The computational resources of an autonomous vehicle can be distributed amongst all subsystems. This distribution must be changed as a function of the situation or scenario the vehicle is facing. To do so, different system performance parameters can be evaluated. The resource allocation of the system can then be optimized to give best possible performance for the whole system. The system performance is given as a weighted sum of all performance metrics, calculated across all subsystems.

$$J(C) = \sum_{i=1}^{n} \left\{ w_i \left( \sum_{j=1}^{m} Z_j \times \sigma_j \right) \right\} \quad (1)$$

This gives the total system performance measurement for $n$ subsystems, each evaluated for $m$ performance metrics. The multiplier $Z_j$ is 1 when a lower value of performance metric is desired and -1 when a higher value of performance matric is desired. Minimizing J in (1) yields desired computational resource distribution vector C, $C \in \mathbb{R}^n$.

The resource distribution is restricted by practical parameters, safety considerations, or the capacity of the subsystem themselves. These constraints are given as (2).

$$\left. \begin{array}{c} C_{min} \leq C_i \leq C_{max} \\ \sum_{i=1}^{n} C_i = C_{max} \\ C_1 \geq \gamma_1 \\ \vdots \\ C_n \geq \gamma_n \end{array} \right\} \quad (2)$$

### B. System Framework for Adaptive Optimization

The framework used to calculate optimum resource allocation is given in Fig. 1. The system consists of camera sensors amongst which the resources are to be distributed. Depending on the context of the situation of the surrounding environment, a contextual priority assignment is given to each of the 4 systems. This priority index is affected by the distance of activity of interest on that side of the camera system. A system model is derived to estimate the performance of the system upon giving certain inputs such as resources, size of the neural network, etc. The output parameters of the model are prediction time required by each subsystem, precision of detection, and energy consumption while doing so. The priority index and system model are used to create the system objective function. This function is developed to maximize performance and minimize the energy consumption of the system. The objective function is optimized over certain constraints that are decided by the limitations on resource availability and lower bounds on allocation. The performance of the system is then checked for safety metrics. The safety metric is evaluated based on the situation and the respective performance obtained by optimum resource allocation.

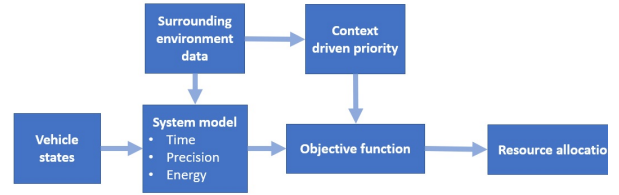

Fig. 1. Framework of adaptive optimization

### C. Modeling of Computational Factors

The performance metrics are calculated as a function of the computational resources, size of the neural network, and the camera resolution of each subsystem. Different models used for time, precision, and energy calculation of each subsystem are given below.

#### 1) Time Consuming Model

As discussed earlier, the processing time for each system depends on the resources allocated to that system, the size of the neural network used, and the resolution of the images. It is assumed that a simple relation can be made to estimate the time required. The time required by the $i^{th}$ system is taken as a function of these parameters as given in (3), where $C_i$, $S_i$, and $R_i$ are the normalized resources allocated, normalized neural network size, and camera resolution of the $i^{th}$ subsystem, respectively. It is assumed that the time required to analyze one image is directly proportional to the resolution of the image, inversely proportional to the size of the neural network used along with the resources available to the system.

$$t_i = f_t(C_i, S_i, R_i) = t_0 + t_1 \times \left( \frac{1}{C_i} \times S_i \times R_i \right) \quad (3)$$

#### 2) Precision Model

The precision of image processing depends on the size of the neural network used and the resolution of the input image. It is not affected by the computational resource distribution. Fewer resources will take longer time to analyze the same information and vice versa. The precision model used for the $i^{th}$ system is given in (4). This relation is also assumed to be proportional with resolution, i.e., higher resolution gives better accuracy. The relation with the size of the neural network is also assumed to be proportional.

$$P_i = f_p(S_i, R_i) = p_0 + p_1 \times S_i \times R_i \quad (4)$$

The image resolution fed to the neural network system changes as a function of importance factor $\alpha_i$. In our framework, multiple artificial neural networks with different image sizes and network dimensions are trained for one given perception task. The minimum image resolution is given as $R_{min}$ and the maximum resolution is given as $R_{max}$ are constant resolution values. The image resolution of the $i^{th}$ system is given by (5).

$$R_i = max\{R_{max} \times \alpha_i, R_{min}\} \quad (5)$$

The size of neural network changes as a function of weightages given by (8) between a minimum normalized value of $S_{min}$ to $S_{max}$. System with higher weightage value gets larger neural network size. The size of neural network of the $i^{th}$ system is calculated by (6).

$$S_i = max\{S_{max} \times W_i, S_{min}\} \quad (6)$$

### 3) Energy Consumption Model

The energy required by each system is modeled as a function of the resources used by it and the frequency of detection of each subsystem. The frequency of detection is the number of times a subsystem captures and analyzes visual information per second. It is changed based on the context of the situation as well as the speed of the vehicle. High speed and visually critical scenario demand for higher detection frequency while at low speeds and situations with low emphasis on surroundings can use a low detection frequency for a subsystem. The energy consumption of the $i^{th}$ subsystem per second is given by (7), where $F_i$ is the frequency of detection. The energy increases linearly with detection frequency, and it is proportional to computational resources being used by the corresponding system.

$$E_i = f_e(C_i, F_i) = e_0 + e_1 \times C_i \times F_i \quad (7)$$

### D. Context-driven Priority Assignment

The priority of each subsystem is determined by the context of the situation the vehicle is facing. The subsystem that is the most critical from the situational perspective gets the highest priority. This priority is given as weightages $W_i$. The weightage given for each system depends on the activity of interest related to each subsystem. The weightage is limited to a minimum value $\alpha_{min}$ so that each subsystem gets minimum importance even when there is no benefit in overall performance. This is an important consideration for safety. The weightage then calculated considering the importance factor of the $i^{th}$ system is given by (8).

$$W_i = \frac{\alpha_i}{\sum_{j=1}^{n} \alpha_j} \quad (8)$$

The frequency of detection is determined based on the context of the situation as well as the speed of the vehicle given by (9). $\alpha_i$ is the importance factor determined from the situation, $v$ is the speed of the vehicle and $v_{f_{max}}$ is the speed of the vehicle above which maximum detection frequency is required. $F_{max}$ is the maximum detection frequency of a subsystem, and $F_{min}$ is the minimum detection frequency of a subsystem. The detection frequency is high when the vehicle speed is high and also when the importance factor of the system is high.

$$F_i = f_f(\alpha_i, v) = (F_{max} - F_{min}) \times \frac{\alpha_i - \alpha_{min}}{1 - \alpha_{min}} \times \frac{v}{v_{f_{max}}} + F_{min} \quad (9)$$

### E. Adaptive Optimization for Computational Resource Allocation

The flow of adaptive resource optimization can be represented by the process given below. The perception system obtains surrounding data and analyzes each subsystem for its importance. A priority index is then assigned to each subsystem. Final system weightage is determined by comparing the relative importance and minimum importance factors of all systems. Depending on the criticality of situation, maximum resources to be distributed, $C_{max}$, is determined. It changes based on the demand of the situation. For a demanding situation, more resources can be used while for a situation where the need for computational power is low, the total resources can be limited to a lower value. The importance factor $\alpha$ of each system can be used to determine the nature of the situation. The resources available for distribution are given in (10).

$$C_{max} = C_0 + C_1 * \sum_{i=1}^{n} \alpha_i \quad (10)$$

The resource distribution is calculated based on a cost function that consists of all the three performance metrics, i.e. time, precision, and energy consumption along with the priority index that is calculated as explained earlier. The priority index calculated is used with the precision metric of each subsystem as it reflects the relative emphasis of each subsystem. The cost function is developed to minimize time as well as energy consumption and maximize the precision of the vehicle. The cost function is given as (11).

$$J = \sum_{i=1}^{n} \{a \times f_t(C_i, S_i, R_i) - b \times f_p(S_i, R_i) W_i + c \times f_e(C_i, F_i)\} \quad (11)$$

where, a, b and c are the weightages that signify the relative importance of cumulative time, system precision, and energy consumption, respectively.

$$\left. \begin{array}{c} C_{min} \leq C_1 \leq C_{max} \\ \vdots \\ C_{min} \leq C_n \leq C_{max} \\ \sum_{i=1}^{n} C_i = C_{max} \\ a + b + c = 1 \\ SF \geq SF_{CCRA} \end{array} \right\} \quad (12)$$

The minimization is constrained by the total resources distributed, minimum resources per system, weightages of performance parameters, and operational safety factor $SF_{CCRA}$ as given in (12).

The resource distribution algorithm works as given in Algorithm 1. The vehicle receives information about surrounding traffic objects and other vehicles from sensors. Based on this information and vehicle speed, the importance factors $\alpha_i$ are calculated for each side. Then the weightage factors $w_i$ are calculated from $\alpha_i$. The detection frequency for each sensor is determined based on importance and speed of vehicle. Then maximum resources available for allocation are determined. Camera resolution and neural network size are determined based on their importance factor and weightage. The objective function is then optimized for the best system performance. Resource allocation is obtained as an optimization variable.

**Algorithm 1: Adaptive Optimization**

Let $V_{rev}$ be the Vehicle direction Boolean (1 for reverse, 0 for forward)
Let $v$ be the speed of the vehicle
Let $D_l$ be the distance of the vehicle on the left side.
Let $D_r$ be the distance of the vehicle on the right side.
Let $D_f$ be the distance of the vehicle ahead.
Let $N$ be the number of subsystems for resource distribution
**if** $V_{rev}$ **then**
   set front importance factor, $\alpha_f = 0$,
   set rear importance factor, $\alpha_b = 1$,
**else**
   set front importance factor, $\alpha_f = 1$
   set rear importance factor, $\alpha_b = 0$
**end**
calculate left importance factor from $D_l$, $\alpha_l = f_{imp}(D_l)$
calculate right importance factor from $D_r$, $\alpha_r = f_{imp}(D_r)$
**for** i → 1 to N
   set minimum importance factor $\alpha_{min}$, $\alpha_i = \max(\alpha_i, \alpha_{min})$
**end**
calculate maximum resources, $C_{max} = f_{max\ resources}(\sum \alpha_i, C_{min})$
**for** i → 1 to N
   calculate weightages $w_i = \alpha_i / \sum \alpha_i$
   calculate detection frequency $F_i = f_{detection\ frequency}(\alpha_i, v)$
   calculate camera resolution $R_i = F_{camera\ resolution}(\alpha_i, R_{min})$
   calculate the size of the neural network, $S_i = F_{neural\ network\ size}(w_i, S_{min})$
**end**
minimize objective function, $C_i$ = minimize { $J(C_i, S_i, R_i, F_i)$ }
**return** Optimum computational resource allocation ($C_i$)

## III. RESULTS AND ANALYSIS

### A. Simulation Setup

For the simulation purpose, a system with 4 camera sensors is in use. Each camera system provides visual input to the perception system and consumes some of the computational resources available. Fig. 2 shows an example of a traffic scenario that the vehicle might be facing. Such a scenario includes traffic objects such as vehicles, pedestrians, or bicyclists in the area surrounding the vehicle. The vehicle receives visual data from onboard camera systems and gets information about its surroundings.

*1) System Models*

The calculation time of subsystems varies from 10 to 50 milliseconds. Constants used for the time model, precision model, and energy model are listed in Table 1.

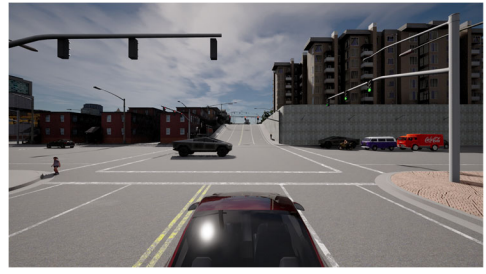

Fig. 2. Traffic scenario example

TABLE 1
SIMULATION CONSTANTS

| Constant | | Value |
|---|---|---|
| Time model | $t_0(s)$ | 0.005 |
| | $t_1(s/MP)$ | 0.002 |
| Precision model | $p_0$ | 0.005 |
| | $p_1(1/MP)$ | 0.3618 |
| Energy model | $e_0$ | 0.005 |
| | $e_1(s)$ | 0.049 |
| Minimum detection frequency | $F_{min}(Hz)$ | 5 |
| Maximum detection frequency | $F_{max}(Hz)$ | 20 |
| Maximum detection frequency speed | $v_{f_{max}}(mph)$ | 40 |
| Minimum image resolution | $R_{max}(MP)$ | 5 |
| Maximum image resolution | $R_{min}(MP)$ | 3 |
| Size of neural network | $S_{min}$ | 0.15 |
| | $S_{max}$ | 1 |
| Maximum resource calculation | $C_0$ | 0 |
| | $C_1$ | 0.3226 |
| Minimum resource | $C_{min}$ | 0.1 |

MP = Megapixels.

The weightages used for time, precision, and energy consumption in optimization function in (11) are listed in Table 2 along with the weightages used for the time penalty, precision penalty, and frequency penalty respectively for safety metric calculation.

TABLE 2
WEIGHTAGES FOR SIMULATION

| Weightages | | Value |
|---|---|---|
| Time weightage | a | 0.5 |
| Precision weightage | b | 0.25 |
| Energy weightage | c | 0.25 |
| Time penalty weightage | $W_p$ | 0.4 |
| Precision penalty weightage | $W_t$ | 0.4 |
| Frequency penalty weightage | $W_f$ | 0.2 |

*2) Context-driven Priority Assignment*

The priority assignment for 4 camera system is done depending on the traffic situation and vehicle behavior. The importance factors $\alpha_i$ vary from 0 to 1, with 0 signifying no traffic or important activity for that subsystem, and 1 signifying the most critical activity. The importance factors of all subsystems change as a function of the distance of activity of interest from the vehicle. All the importance factors are calculated based on distances given in Fig. 3. If the vehicle is taking a turn on either side, the corresponding subsystem is given an importance factor of 1 regardless of its distance from the vehicle in the left or right lane.

The scenarios used to measure the performance of dynamic resource distribution consist of the distance of the front vehicle, the distance of vehicles in the left and right lane, and the speed of the vehicle. If the vehicle is turning or going in reverse, it is also considered in the scenario. The 14 scenarios given in Table 3 represent a 5-minute journey of a passenger vehicle.

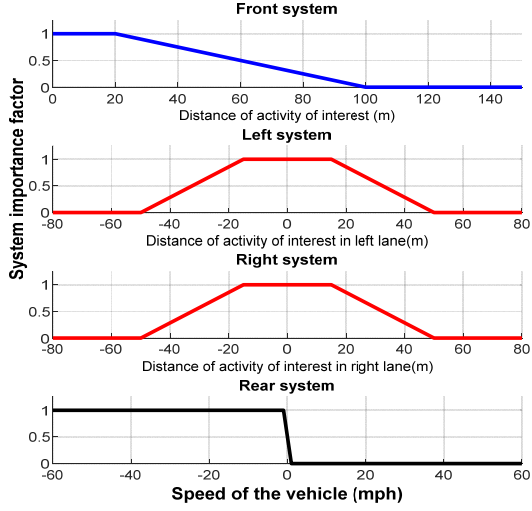

Fig. 3. Contextual importance factors for all subsystems

TABLE 3
DRIVING SCENARIOS

| Scenario No. | Speed (mph) | Direction of motion | Location of activity of interest | Distance (m) |
|---|---|---|---|---|
| 1 | 5 | Reverse and left turn | Left lane | 5 |
|  |  |  | Right lane | 10 |
| 2 | 10 | Straight | Front | 10 |
|  |  |  | Left lane | 10 |
|  |  |  | Right lane | 10 |
| 3 | 10 | Right turn | Front | 50 |
| 4 | 35 | Left lane change | Front | 10 |
| 5 | 45 | Straight | Front | 32 |
| 6 | 9 | Straight | Front | 11 |
|  |  |  | Left lane | 20 |
|  |  |  | Right lane | 20 |
| 7 | 36 | Straight | Front | 10 |
|  |  |  | Left lane | 5 |
| 8 | 45 | Right lane change | Left lane | 20 |
| 9 | 61 | Left lane change | Left lane | 5 |
| 10 | 65 | Straight | Front | 30 |
|  |  |  | Left lane | 42 |
| 11 | 50 | Right lane change | Front | 38 |
| 12 | 45 | Straight | None | - |
| 13 | 10 | Right turn | None | - |
| 14 | 7 | Reverse and right turn | Right lane | 7 |

*3) Operational Safety Metric*

The total resources used by the dynamic resource allocation change with the scenario. Therefore, it is needed to maintain operational safety that ensures no subsystem falls short of resources when the situation is demanding. A safety metric is used to compare the safety of the system in all cases. For the safety calculation, a penalty-based approach is used. When there is a need for a higher emphasis on a particular system out of the 4 systems, a penalty is given for that subsystem if its operational metrics are not high enough. This normalized penalty is a combined measure of precision, calculational time, and the detection frequency. The weighted sum of these metrics gives rise to the penalty of that subsystem. The penalty of the $i^{th}$ system is given by (13). $W_p$, $W_t$, and $W_f$ are the weightages given for precision, time, and frequency penalties. $F_{ideal}$ is the ideal detection frequency calculated from (9).

$$\beta_i = (1 - P_i) \times W_p + (t_i - 0.01) \times T_n \times W_t + \frac{F_{ideal} - F_i}{F_{ideal} - F_{min}} \times W_f \quad (13)$$

The total penalty is calculated by adding penalties of all the subsystems with the weightages given by (8). The initial normalized safety is taken as 1 and the total penalty is subtracted to get safety measures of the system. The total safety metric must always be more than or equal to safety metrics calculated for constant computational resource allocation (CCRA) for any given situation. The final safety for the overall system is calculated as given in (14).

$$SF = 1 - \sum_{i=1}^{4} \beta_i \times W_i \quad (14)$$

*B. Results and Analysis*

The optimal resource distribution is calculated for the 12 scenarios given in Table 3. The optimization problem is solved using the Nelder-Mead simplex algorithm [18]. The performance of the system is evaluated by metrics of maximum detection time, cumulative precision of the system, and energy consumption per second. The safety metric is also compared for its corresponding value with the conventional resource allocation system (CCRA). All the performance metrics are evaluated in terms of normalized values.

The performance of the system with adaptive resource allocation is given in Table 4. To compare the performance of the proposed adaptive computational resource allocation (ACRA) system with CCRA, the performance of a constant resource allocation system should be evaluated in terms of the same performance metrics. It is considered that the resources for a conventional system are distributed equally amongst all subsystems and do not depend on the scenario. The size of the neural network will be equally distributed for each subsystem. The image resolution and the detection frequency used by such a system is always the maximum. Regardless of the scenario, the system performance with CCRA remains constant since there is no change in parameters such as resource allocation, image resolution, or detection frequency. The performance of the system with CCRA is given in the bottom row in Table 4.

The resource distribution with ACRA varies significantly. As the scenario becomes demanding, such as for the case of scenario 2, the total resources available for distribution are high. For a scenario with a lower demanding situation such as scenario 12, the maximum resources available for distribution are significantly lower. The rear visual system gets considerable resources only when the vehicle is moving in reverse as in scenarios 1 or 14. Thus the resources available for other systems are higher. When the same case is compared with CCRA, the reverse visual system has access to equal amount of resources. As a result, the resources are not optimally utilized. The ACRA method is better in allocating resources where the potential utilization is better.

TABLE 4
PERFORMANCE METRICS OF ACRA AND CCRA METHODS

| Scenario No. | Resources allocated (normalized) | | | | | Total time (s) | Max time (s) | Precision (normalized) | | | | | Detection frequency (Hz) | | | | Total energy per second (normalized) | Safety metric (normalized) |
|---|---|---|---|---|---|---|---|---|---|---|---|---|---|---|---|---|---|---|
| | C1 (front) | C2 (left) | C3 (right) | C4 (rear) | Total | | | P1 (front) | P2 (left) | P3 (right) | P4 (rear) | total | F1 (front) | F2 (left) | F3 (right) | F4 (rear) | | |
| 1 | 0.29 | 0.24 | 0.24 | 0.24 | 1 | 0.064 | 0.019 | 0.17 | 0.59 | 0.59 | 0.59 | 1.93 | 5 | 6.88 | 6.88 | 6.88 | 0.33 | 0.82 |
| 2 | 0.18 | 0.18 | 0.18 | 0.45 | 1 | 0.075 | 0.023 | 0.59 | 0.59 | 0.59 | 0.17 | 1.93 | 8.75 | 8.75 | 8.75 | 5 | 0.37 | 0.81 |
| 3 | 0.14 | 0.13 | 0.19 | 0.13 | 0.59 | 0.078 | 0.033 | 0.39 | 0.17 | 1 | 0.17 | 1.72 | 7.19 | 5 | 8.75 | 5 | 0.21 | 0.89 |
| 4 | 0.12 | 0.12 | 0.24 | 0.24 | 0.71 | 0.106 | 0.044 | 0.83 | 0.83 | 0.17 | 0.17 | 1.99 | 18.13 | 18.13 | 5 | 5 | 0.34 | 0.9 |
| 5 | 0.11 | 0.09 | 0.09 | 0.09 | 0.37 | 0.095 | 0.049 | 0.84 | 0.17 | 0.17 | 0.17 | 1.35 | 17.4 | 5 | 5 | 5 | 0.17 | 0.91 |
| 6 | 0.2 | 0.18 | 0.18 | 0.33 | 0.89 | 0.068 | 0.023 | 0.66 | 0.46 | 0.46 | 0.17 | 1.74 | 8.38 | 7.73 | 7.73 | 5 | 0.32 | 0.79 |
| 7 | 0.24 | 0.11 | 0.11 | 0.24 | 0.71 | 0.107 | 0.045 | 0.17 | 0.83 | 0.83 | 0.17 | 1.99 | 5 | 18.5 | 18.5 | 5 | 0.35 | 0.9 |
| 8 | 0.22 | 0.1 | 0.11 | 0.22 | 0.66 | 0.106 | 0.048 | 0.17 | 0.65 | 0.88 | 0.17 | 1.87 | 5 | 17.62 | 20 | 5 | 0.33 | 0.88 |
| 9 | 0.1 | 0.11 | 0.1 | 0.1 | 0.42 | 0.096 | 0.055 | 0.17 | 1 | 0.17 | 0.17 | 1.5 | 5 | 20 | 5 | 5 | 0.2 | 0.93 |
| 10 | 0.11 | 0.1 | 0.18 | 0.18 | 0.58 | 0.096 | 0.043 | 0.78 | 0.52 | 0.17 | 0.17 | 1.64 | 17.92 | 15.24 | 5 | 5 | 0.28 | 0.86 |
| 11 | 0.1 | 0.21 | 0.11 | 0.21 | 0.64 | 0.103 | 0.049 | 0.56 | 0.17 | 0.92 | 0.17 | 1.81 | 16.25 | 5 | 20 | 5 | 0.32 | 0.88 |
| 12 | 0.03 | 0.03 | 0.03 | 0.03 | 0.13 | 0.206 | 0.052 | 0.28 | 0.28 | 0.28 | 0.28 | 1.11 | 5 | 5 | 5 | 5 | 0.05 | 0.93 |
| 13 | 0.09 | 0.09 | 0.16 | 0.09 | 0.42 | 0.086 | 0.039 | 0.17 | 0.17 | 1 | 0.17 | 1.5 | 5 | 5 | 8.75 | 5 | 0.15 | 0.94 |
| 14 | 0.15 | 0.15 | 0.21 | 0.21 | 0.71 | 0.076 | 0.027 | 0.17 | 0.17 | 0.83 | 0.83 | 1.99 | 5 | 5 | 7.63 | 7.63 | 0.25 | 0.91 |
| CCRA | 0.25 | 0.25 | 0.25 | 0.25 | 1 | 0.06 | 0.015 | 0.457 | 0.457 | 0.457 | 0.457 | 1.829 | 20 | 20 | 20 | 20 | 1 | 0.778 |

The total resources available for ACRA change depending on the context of situation and complexity. For critical situations, higher resources are available. For relatively non-demanding situations, the total resources available for distribution are lower. Thus, system performance is maintained by utilizing resources only when they are necessary. With CCRA, the resources are always used completely. Thus, underutilization of resources takes place when there is no need for all the resources to be allocated.

The detection time of each subsystem depends on the computational resources allocated, size of neural network, and camera resolution. As the resources allocated to the system with lower importance are low, the detection time for these systems is high. For systems with higher contextual priority, the detection time is slightly higher than that with CCRA as a result of lower resources and image resolution. However, this difference is since the conventional system always runs with all available resources. In Fig. 4, the total time of detection for all subsystems combined is given. The total time for the ACRA method is higher than that for the CCRA method. However, this is not always favorable since, for lower demanding situations, the detection time requirement is not critical. In such a situation, the performance of subsystems that are relevant to the situational context plays a role in total system performance.

In table 4, the precision for ACRA system is higher than CCRA for most of the scenarios. For a system that is important for the given situation, the precision obtained is always higher than the respective precision with CCRA method. Take the example of scenario 8. Here, the vehicle makes a right lane change. It is thus important to improve the precision of the right camera system. The results ACRA method gives higher precision for this system as compared to the CCRA method. For conventional system, the precision is low as a result of relatively low resources. It can be observed that the precision for subsystems that are not relevant from a contextual point have lower precision as they do not contribute significantly to the total performance of the system.

The detection frequency for CCRA is always at the highest while the detection frequency for the ACRA method changes as the scenario around the vehicle changes. For scenarios 9 and 10, the vehicle is traveling at high speed. The detection frequency of subsystems that are critical for a given scenario is maintained at high values. The subsystems that do not play a significant role in the situation operate at lower detection frequency. Thus, higher detection frequency is used only when the situation is demanding and/or at high vehicle speeds. The computational load on the visual system is thus reduced and unnecessary usage of computational resources is avoided. Lower detection frequency also improves energy usage with the ACRA system.

Energy consumed by all the subsystems depends on the frequency of detection and the resources allocated to each system along with total resources available. In Fig. 4 it can be seen that the energy consumed by the adaptive resource allocation system is always lower than the energy consumed by vehicle with CCRA. This is because, with the conventional system, all the subsystems operate at the maximum possible frequency of detection, and all the available computational resources are always consumed.

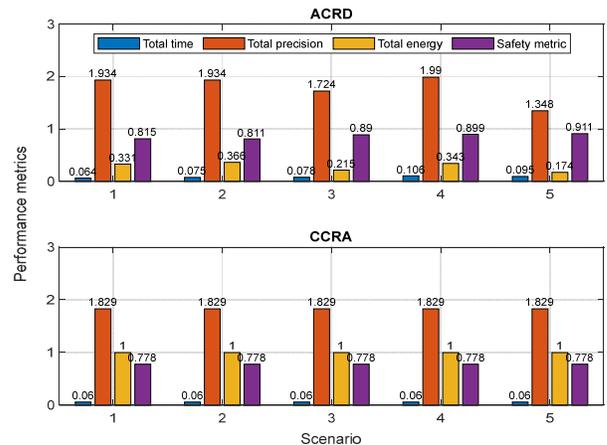

Fig. 4. Comparison of performance metrics

The safety metric with ACRA is always higher than that for CCRA. This is very important from an operational point of view. The safety of the vehicle is improved with higher performance in other aspects. The safety of the vehicle depends on the performance metrics of the subsystems that are critical for a given context. With ACRA, the performance of such subsystems is always better. Thus, the safety metric is always higher. The safety with the conventional system is restricted by the equal allocation of resources. The subsystems that are critical for a given situation receive equal resources as other systems and thus get restricted in performance.

## IV. Conclusions

The proposed method for the distribution of on-board computational resources provides superior performance as compared to the conventional method. The total resources available for distribution change as per scenario and optimize the system performance. A subsystem that is critical for a given scenario receives higher resources and thus gives better performance. The ACRA method allocates resources where their utilization gives optimal system performance and lower energy consumption. The precision for the proposed approach varies depending on the complexity of the scenario. It results in a tradeoff between overall precision for lower energy consumption when compared with conventional methods of resource allocation. The dynamic detection frequency results in lower computational load and energy consumption for non-demanding situations. With ACRA, vehicle performance improves for critical situations, and the safety metric is always maintained higher than conventional methods of resource allocation. The ACRA method can optimize the resource allocation of autonomous vehicles with improved performance, safety, and energy consumption.

## References

[1] J. O. Kephart et al., "Coordinating multiple autonomic managers to achieve specified power-performance tradeoffs," 2007, doi: 10.1109/ICAC.2007.12.

[2] K. C. Gouda, T. V Radhika, M. Akshatha, and others, "Priority based resource allocation model for cloud computing," *Int. J. Sci. Eng. Technol. Res.*, vol. 2, no. 1, pp. 215–219, 2013.

[3] M. Wang, N. Kandasamy, A. Guez, and M. Kam, "Adaptive performance control of computing systems via distributed cooperative control: Application to power management in computing clusters," in *Proceedings - 3rd International Conference on Autonomic Computing, ICAC 2006*, 2006, vol. 2006, pp. 165–174, doi: 10.1109/icac.2006.1662395.

[4] R. P. Doyle, J. S. Chase, O. M. Asad, W. Jin, and A. Vahdat, "Model-Based Resource Provisioning in a Web Service Utility.," in *USENIX Symposium on Internet Technologies and Systems*, 2003, vol. 4, p. 5.

[5] M. Thammawichai and E. C. Kerrigan, "Energy-efficient real-time scheduling for two-type heterogeneous multiprocessors," *Real-Time Syst.*, vol. 54, no. 1, pp. 132–165, Jan. 2018, doi: 10.1007/s11241-017-9291-6.

[6] J. Sun, Q. Gu, T. Zheng, P. Dong, and Y. Qin, "Joint communication and computing resource allocation in vehicular edge computing," *Int. J. Distrib. Sens. Networks*, vol. 15, no. 3, p. 1550147719837859, 2019.

[7] S. Goudarzi, M. H. Anisi, H. Ahmadi, and L. Musavian, "Dynamic Resource Allocation Model for Distribution Operations Using SDN," *IEEE Internet Things J.*, vol. 8, no. 2, pp. 976–988, Jan. 2021, doi: 10.1109/JIOT.2020.3010700.

[8] J. Zhang, W. Xia, F. Yan, and L. Shen, "Joint computation offloading and resource allocation optimization in heterogeneous networks with mobile edge computing," *IEEE Access*, vol. 6, pp. 19324–19337, 2018.

[9] Y. Zhan, P. Li, and S. Guo, "Experience-Driven Computational Resource Allocation of Federated Learning by Deep Reinforcement Learning," in *Proceedings - 2020 IEEE 34th International Parallel and Distributed Processing Symposium, IPDPS 2020*, May 2020, pp. 234–243, doi: 10.1109/IPDPS47924.2020.00033.

[10] L. Liang, H. Ye, G. Yu, and G. Y. Li, "Deep-Learning-Based Wireless Resource Allocation with Application to Vehicular Networks," *Proc. IEEE*, vol. 108, no. 2, pp. 341–356, Feb. 2020, doi: 10.1109/JPROC.2019.2957798.

[11] A. Huang, Y. Li, Y. Xiao, X. Ge, S. Sun, and H. C. Chao, "Distributed Resource Allocation for Network Slicing of Bandwidth and Computational Resource," in *IEEE International Conference on Communications*, Jun. 2020, vol. 2020-June, doi: 10.1109/ICC40277.2020.9149296.

[12] G. Sun, G. O. Boateng, D. Ayepah-Mensah, G. Liu, and J. Wei, "Autonomous Resource Slicing for Virtualized Vehicular Networks with D2D Communications Based on Deep Reinforcement Learning," *IEEE Syst. J.*, vol. 14, no. 4, pp. 4694–4705, Dec. 2020, doi: 10.1109/JSYST.2020.2982857.

[13] S. Matzka, A. M. Wallace, and Y. R. Petillot, "Efficient resource allocation for attentive automotive vision systems," *IEEE Trans. Intell. Transp. Syst.*, vol. 13, no. 2, pp. 859–872, 2012, doi: 10.1109/TITS.2011.2182610.

[14] P. Dziurzanski, A. K. Singh, and L. S. Indrusiak, "Energy-aware resource allocation in multi-mode automotive applications with hard real-time constraints," in *Proceedings - 2016 IEEE 19th International Symposium on Real-Time Distributed Computing, ISORC 2016*, Jul. 2016, pp. 100–107, doi: 10.1109/ISORC.2016.23.

[15] S. Baidya, Y. J. Ku, H. Zhao, J. Zhao, and S. Dey, "Vehicular and edge computing for emerging connected and autonomous vehicle applications," in *Proceedings - Design Automation Conference*, Jul. 2020, vol. 2020-July, doi: 10.1109/DAC18072.2020.9218618.

[16] H. Zhao et al., "Towards safety-aware computing system design in autonomous vehicles," *arXiv Prepr. arXiv1905.08453*, 2019.

[17] J. Wang, C. Huang, K. He, X. Wang, X. Chen, and K. Qin, "An energy-aware resource allocation heuristics for VM scheduling in cloud," in *Proceedings - 2013 IEEE International Conference on High Performance Computing and Communications, HPCC 2013 and 2013 IEEE International Conference on Embedded and Ubiquitous Computing, EUC 2013*, 2014, pp. 587–594, doi: 10.1109/HPCC.and.EUC.2013.89.

[18] J. C. Lagarias, J. A. Reeds, M. H. Wright, and P. E. Wright, "Convergence properties of the Nelder--Mead simplex method in low dimensions," *SIAM J. Optim.*, vol. 9, no. 1, pp. 112–147, 1998.